\documentclass[preprints,article,accept,moreauthors,pdftex]{Definitions/mdpi} 

\firstpage{1} 
\pubvolume{1}
\issuenum{1}
\articlenumber{0}
\pubyear{2021}
\copyrightyear{2020}
\datereceived{} 
\dateaccepted{} 
\datepublished{} 
\hreflink{https://doi.org/} 

\newcommand{\beq}{\begin{equation}}
\newcommand{\eeq}{\end{equation}}
\newcommand{\bea}{\begin{eqnarray}}
\newcommand{\eea}{\end{eqnarray}}
\newcommand{\tr}{\text{Tr}}

\usepackage{physics}
\usepackage{bbm}

\extrafloats{1000}

\Title{Toward an automated-algebra framework for high orders in the virial expansion of quantum matter}

\TitleCitation{Title}

\Author{Aleks J. Czejdo$^{1,\ddagger,}$\orcidA{}, Joaquin E. Drut $^{2,\ddagger,}$*\orcidA{}, Yaqi Hou $^{3,\ddagger}$\orcidC{} and Kaitlyn J. Morrell$^{4,\ddagger}$\orcidD{}}

\AuthorNames{Aleks J. Czejdo, Joaquin E. Drut, Yaqi Hou and Kaitlyn J. Morrell}

\AuthorCitation{Czejdo, A. J.; Drut, J. E.; Hou, Y.; Morrell, K. J.}

\address[1]{%
$^{\ddagger}$ \quad Department of Physics and Astronomy, University of North Carolina, Chapel Hill, North Carolina 27599, USA\\
$^{1}$ \quad aczejdo@live.unc.edu\\
$^{2}$ \quad drut@email.unc.edu\\
$^{3}$ \quad yaqi.hou@unc.edu\\
$^{4}$ \quad nyltiak@live.unc.edu\\}

\corres{Correspondence: drut@email.unc.edu}

\abstract{ The virial expansion provides a non-perturbative view into the
  thermodynamics of quantum many-body systems in dilute regimes. While
  powerful, the expansion is challenging as calculating its coefficients at
  each order $n$ requires analyzing (if not solving) the quantum $n$-body
  problem.  In this work we present a comprehensive review of automated
  algebra methods, which we developed to calculate high-order virial
  coefficients.  The methods are computational but non-stochastic, thus
  avoiding statistical effects; they are also for the most part analytic,
  not numerical, and amenable to massively parallel computer architectures.
  We show formalism and results for coefficients characterizing the
  thermodynamics (pressure, density, energy, static susceptibilities) of
  homogeneous and harmonically trapped systems, and explain how to
  generalize them to other observables such as the momentum distribution,
  Tan's contact, and the structure factor.}


\begin{document}
\tableofcontents
\section{Introduction}

Quantum many-body systems are notoriously difficult to compute.
Whenever interactions play an important role, in atomic, condensed matter, or nuclear physics, 
most analytic approaches (if not all) are unable to give quantitatively reliable results and 
numerical methods often face challenges of their own as well, such as the infamous sign problem (notwithstanding remarkable progress over the last few years)~\cite{BERGER20211}.

The virial expansion (VE) (see e.g. Ref.~\cite{pathria1972statistical} for an introduction to
both the classical and quantum cases and Ref.~\cite{Liu2013PR} for a comprehensive review) 
aims to tackle the finite-temperature quantum many-body problem
by breaking it down into contributions from subspaces of the full Fock space 
corresponding to fixed (and small) particle number.
In this sense, the VE is effectively an expansion around a dilute limit in which
the interparticle distance is much larger than every other scale in the system, in particular
the thermal wavelength $\lambda_T = \sqrt{2\pi \beta}$, where $\beta = 1/T$ is the
inverse temperature and we have used units such that $\hbar = k_B = m = 1$. 
One might expect that, this being a high-temperature regime (as $\lambda_T$ is in the above 
sense small), interaction effects would play a quantitatively minor role; 
this, however, is not necessarily the case, as we will see. Another misconception is that 
quantum effects play a small role in such a regime, which is also not generally true.
Both interaction and quantum effects are central in the calculation of the virial expansion and
leave a clear imprint in the expansion coefficients. It is for that reason that such calculations 
are challenging and that specialized techniques are required to carry them out, as we explain in this review. But, we are getting ahead of ourselves; let us start from the beginning.

The origin of the VE is the classical cluster expansion of 
Mayer et al, developed in the 1930s~\cite{MayerMayer}, 
in which the classical grand-canonical partition function $\mathcal Z$ of a three-dimensional 
gas is expanded in powers of the fugacity $z = \exp(\beta \mu)$, where $\mu$ is the 
chemical potential, namely 
\beq
\mathcal Z = \sum_{N=0}^{\infty}\left(\frac{z}{\lambda_T^3} \right)^N \frac{Q_N}{N!},
\eeq
where
\beq
Q_N = \int \prod_{i=1}^N d {\bf r}_i e^{-\frac{\beta}{2} \sum_{i\neq j} V_{ij} },
\eeq
is the classical canonical partition function and $V_{ij}$ is the pairwise interaction potential
between particle $i$ and particle $j$. 

An expansion of both the pressure $P$ and the number density $n$ 
in powers of $z$ is thus obtained, namely
\bea
\beta P &=& \frac{1}{\lambda_T^3} \sum_{k = 1}^\infty b_k z^k, \\
n &=& \frac{1}{\lambda_T^3} \sum_{k = 1}^\infty k b_k z^k,
\eea
where the $b_k$, originally called `cluster coefficients', typically depend on
temperature and the specific form of $V_{ij}$.
Substituting the density expansion into the pressure expansion, order by order in $z$, one obtains
the old virial expansion of the imperfect gas equation of state in powers of the density, i.e.
\beq
P = n k_B T \sum_{k=1}^{\infty} a_k x^{k-1},
\eeq
where $x = n \lambda_T^3$ and the $a_k$ are in older literature called `virial coefficients' 
(more recently, that nomenclature has been used for the $b_k$ coefficients instead).
Calculating the $a_k$ requires knowing all the $b$ coefficients up to order $k$, which in 
turn requires calculating the canonical partition functions of up to
$k$ particles, as we will show in more detail below.
The original work of Mayer et al proposed a diagrammatic technique for calculating the classical case, which is naturally simpler than its quantum counterpart simply because of the well known fact that kinetic and potential energies are commuting numbers in the classical case and non-commuting operators in the quantum case. The quantum virial expansion was first explored
by Kahn and Uhlenbeck~\cite{KahnUhlenbeck} and further developed by 
Lee and Yang~\cite{LeeYang1959a,LeeYang1959b,LeeYang1960a,LeeYang1960b,LeeYang1960c}.
As we review below, the calculation of $b_k$ for the quantum case is so challenging in practice that efforts to calculate the third-order coefficient and beyond were not successful until the 
21st century, when computers became powerful enough to apply exact diagonalization techniques. From this point on, we focus entirely on the VE in the context of quantum systems.

Historically, applications of the VE have followed the above paradigm and thus centered on 
equations of state. However, since the VE is an expansion of the grand-canonical partition 
function, it is possible to apply it to any physical quantity. We will show 
techniques to calculate the VE for applications to pressure and density but also to the Tan 
contact~\cite{Tan2008AoP,Tan2008AoPa,Tan2008AoPb} (relevant for systems with short-range interactions), the momentum distribution, and response functions such as the compressibility and the structure factor.

The recent developments of automated algebra, led by our group
~\cite{PhysRevA.98.053615, PhysRevA.100.063626, Hou2019PRA, Czejdo2020PRA, Hou2020PRL, Hou2020PRA, Hou2021PRRFourth, Rammelmueller2021PRA}, have enabled the
precise calculation of high-order coefficients (meaning beyond the third order, which can currently
be addressed numerically). With such orders in hand, it becomes practical and meaningful to 
implement resummation techniques which, uncertainties notwithstanding, have been shown to 
substantially extend the domain of applicability of the virial expansion~\cite{Hou2020PRL, Hou2020PRA}. 

The remainder of this work is organized as follows. 
We begin in Sec.~\ref{sec:background} by reviewing 
the formal elements of the virial expansion of the grand thermodynamic potential, as that is 
the simplest case, which also allows us to establish the basic identities and notation.
In Sec.~\ref{sec:brief-review}, we provide a brief review of conventional calculation methods
for the virial coefficients.
In Sec.~\ref{sec:base-case}, we present our method by example using a
homogeneous system of Fermion gases with attractive interaction changing
from non-interacting limit to the unitary limit.
Encouraged by the agreements with and improvements over existing results,
we further generalize the method to other systems in
Sec.~\ref{sec:gener-other-syst-1}, including harmonically trapped systems
in Sec.~\ref{sec:harm-trapp-syst}, neutron matter in
Sec.~\ref{sec:neutron-matter}, and the unitary Bose gas in Sec.~\ref{sec:unitary-bose-gas}.
Finally, Sec.~\ref{sec:moment-distr} demonstrates the applications to more
complicated observables. Namely, one-body operators such as density or
momentum distributions are discussed in Sec.~\ref{sec:moment-distr}, and
two-body operators, which are of interest to quantities such as the
structure factor or viscosity, are shown in Sec.~\ref{sec:dens-dens-corr}.

\section{Basic formalism}
\label{sec:background}

As mentioned above, the VE is an expansion in
powers of the fugacity $z$
\beq
z = e^{\beta \mu},
\eeq
where $\beta$ is the inverse temperature and $\mu$ is the chemical potential.
In the presence of spin or other internal degrees of freedom, there will naturally be
a fugacity attached to each (conserved) particle number. In this section we
present the formalism for a single flavor for simplicity, but in upcoming sections
we will generalize it to spin-$1/2$ fermions.

The thermodynamics is encoded in the grand-canonical partition function,
which for a quantum system is given by
\beq
\mathcal Z = \Tr[e^{- \beta (\hat{H} - \mu \hat{N})}],
\eeq
where $\hat H$ is the Hamiltonian, $\hat N$ is the particle number operator, 
and \( \mu \) is the chemical potential.
Expanding $\mathcal Z$ in powers of the fugacity we obtain
\beq
\label{eq:cal-Z-expansion}
\mathcal Z = \sum^{\infty}_{N=0} z^N Q_{N},
\eeq
where 
\beq
Q_N = \tr^{}_N\left [e^{-\beta \hat{H}} \right],
\eeq
is the canonical $N$-particle partition function.

Then, the grand thermodynamic potential is expanded in powers of $z$ 
using the above expressions to obtain the conventional expression
\beq
\label{eq:ln-cal-Z-expansion}
- \beta \Omega = \beta PV = \ln \mathcal{Z} = Q_1 \sum_{k=1}^{\infty} z^k b_k,
\eeq
where \( b_k \) is the \( k \)-th order virial
coefficient, which is an intensive, dimensionless quantity. 
Through Eq.~\eqref{eq:cal-Z-expansion} and
Eq.~\eqref{eq:ln-cal-Z-expansion}, the \( b_k \) are related to the canonical
partition functions; for example, the first few \( b_k \) are

\begin{equation}
  \begin{aligned}
    b_1 = &\ 1, \\
    b_2 = & \frac{Q_2}{Q_1} - \frac{Q_1}{2 !}, \\
    b_3 = & \frac{Q_3}{Q_1} - b_2 Q_1 - \frac{Q_1^2}{3!}, \\
    b_4 = & \frac{Q_4}{Q_1} - \left(b_3 + \frac{b_2^2}{2}\right) Q_1 - b_2 \frac{Q_1^2}{2!} - \frac{Q_1^3}{4!}.
  \end{aligned}
\end{equation}
For a system of non-interacting, non-relativistic fermions in \( d \) dimensions, the 
virial coefficients are
\begin{equation}
  \label{eq:1}
  b_k^{(0)} = (-1)^{k+1} k^{- (d + 2) / 2},
\end{equation}
whereas for bosons the result is simply $k^{- (d + 2) / 2}$.
%
%
In the presence of interactions, one usually expands the ratio of $\mathcal Z$ to 
its noninteracting counterpart $\mathcal Z_0$, to obtain
\beq
\label{eq:ln-cal-Z-Z0-expansion}
- \beta \Delta \Omega = \ln (\mathcal{Z}/\mathcal{Z}_0) = Q_1 \sum_{k=1}^{\infty} z^k \Delta b_k,
\eeq
where $\Delta b_k = b_k - b_k^{(0)}$ captures the interaction effects and is the quantity
most often reported.
In practice, the $\Delta b_k$ coefficients are determined by the interaction-induced 
change in $Q_j$ for 
$1 < j \leq k$, and therein lies the difficulty: those $Q_j$ must be determined with
enough accuracy to cancel out all the volume dependence (which contains terms that 
scale with power up to $j$) and obtain a volume-independent $\Delta b_k$.
As we will see below, some methods focus on extracting $\Delta b_k$ directly from
grand-canonical quantities (typically the density), where the volume cancellations have
already happened, while others such as ours (and similarly exact diagonalization) propose 
to calculate the interaction effects on $Q_j$ and use those to calculate $\Delta b_k$.

The above is the VE as applied to the pressure; from it, the expansion for the density
is easily derived. As mentioned above, the VE can in fact be applied to any observable 
such as Tan's contact, the momentum distribution, and the structure factor (see Sec.~\ref{sec:dens-dens-corr}). We will return to those in a later section.

%

\section{Calculation methods for the virial expansion}
\label{sec:brief-review}

\subsection{Second order}

Besides the trivial one-body contribution, fully captured by $Q_1$ and
factored out of the expansion, the leading contribution accounting for 
interaction effects appears at second order, i.e. the two-particle subspace. 
The interaction effects on the two-particle spectrum are captured 
by the scattering properties (binding energies and phase shifts) and in those
terms the second order virial coefficient \( b_2 \) was first calculated analytically by
Beth and Uhlenbeck in the 1930's \cite{Uhlenbeck1936Pquantum, Beth1937Pquantum}.
Specifically, their result relates the interaction change \( \Delta b_2 \) to the
two-body scattering phase shift \( \delta (E) \)
such that, for a spin-1/2 Fermi gas in three spatial dimensions (3D),
\begin{equation}
  \label{eq:3}
\Delta b_2 = \sqrt{2}\sum_i e^{- \beta E_B^i} + \sqrt{2}\sum_l \frac{2 l  + 1}{\pi} \int_0^{\infty} \dd p \dv{\delta_l}{p} e^{- \frac{\lambda_T^2 p^2}{2 \pi}}
\end{equation} 
where the first summation is over all bound states and the second summation over all partial waves.

One may take the above expression in different dimensions and relate it to the
parameters of the corresponding effective range expansion (i.e. scattering length, 
effective range, etc.) and obtain, for a zero-range interaction 
(see e.g.~\cite{Hoffman1D, LoheacDrut2017} for results in 1D,
~\cite{ChafinSchaefer, PhysRevA.97.033630} for 2D, 
and 
~\cite{PhysRevC.73.015201} for 3D):
\begin{equation}
  \label{eq:4}
  \begin{aligned}
  \Delta b_2^\text{1D} &= - \frac{1}{2 \sqrt{2}} + \frac{e^{\lambda_1^2 / 4}}{2 \sqrt{2}} \left[ 1 + \erf(\lambda_1 / 2) \right],\\
  \Delta b_2^\text{2D} &= e^{\lambda_2^2} - 2 \int_0^{\infty} \frac{\dd p}{p} \frac{2 e^{- \lambda_2^2 p^2}}{\pi^2 + 4 \ln^2(p^2) },  \\
  \Delta b_2^\text{3D} &= \frac{e^{\lambda_3^2}}{\sqrt{2}} \left[ 1 + \erf(\lambda_3) \right],  \\
  \end{aligned}
\end{equation}
where \( \lambda_d \) is the physical coupling strength in \( d \) dimensions, defined as
\begin{equation}
  \label{eq:5}
  \begin{aligned}
    \lambda_1 &= 2 \frac{\sqrt{\beta}}{a_0}, \\
    \lambda_2 &= \sqrt{\beta E_B},\\
    \lambda_3 &= \frac{\sqrt{\beta}}{a_0}, \\
  \end{aligned}
\end{equation}
where \( a_0 \) is the s-wave scattering length and $E_B$ is the binding energy
of the single two-body bound state of the 2D case.

While the above results are sufficient for simple systems such as dilute gases of 
ultracold atoms, where the interaction can be modeled very precisely as being
purely zero-range $s$-wave, a deeper analysis is needed to account for the
complexities of nuclear systems (such as neutron and nuclear matter).
%
References~\cite{HOROWITZ2006153, HOROWITZ200655} presented such an extension of the 3D case
to the richer scattering properties found in nuclear physics. There, one must account for
not only finite range but also angular momentum channels beyond the simplest case of pure 
s-wave. For pure neutron matter, Refs.~\cite{HOROWITZ2006153, HOROWITZ200655} integrate by parts to rewrite the Beth-Uhlenbeck result as
\beq
\Delta b_2 = \frac{1}{2^{1/2}\pi T} \int_0^{\infty} dE\ e^{-\beta E/2} \delta^\text{tot}_\text{neutrons}(E),
\eeq
where $\delta^\text{tot}_\text{neutrons}(E)$ is the sum of all the scattering phase shifts at laboratory energy $E$,
whose contributions from different angular momentum channels enter as
\beq
\delta^\text{tot}_\text{neutrons}(E) = \sum_{S,L,J} (2 J +1) \delta_{  ^{2S+1} L_J }(E),
\eeq
where the partial wave terms $\delta_{  ^{2S+1} L_J }(E)$ are obtained from partial wave
analyses of experimental data such as Nijmegen's~\cite{PhysRevC.48.792}.

For nuclear matter, on the other hand, one must account for the deuteron bound state, such that
\beq
\Delta b_2 = \frac{3}{2^{1/2}}\left(e^{E_d/T} - 1\right)   + \frac{1}{2^{3/2}\pi T} \int_0^{\infty} dE\ e^{-\beta E/2} \delta^\text{tot}_\text{nuc}(E),
\eeq
where $E_d$ is the binding energy of the deuteron and the $-1$ term comes from partial 
integration when accounting for the phase shift at zero energy being $\pi$ times the number of 
bound states (see also Ref.~\cite{PhysRevA.100.062110}). The work of Refs.~\cite{HOROWITZ2006153, HOROWITZ200655} also analyzed the contributions due to 
pure alpha-particle scattering and nucleon-alpha scattering, thus obtaining all possible 
contributions to the second-order virial expansion for nuclear matter composed of neutrons, 
protons, and alpha particles.
%

\subsection{Third order and beyond}

The complexity of the quantum many-body problem for three particles and beyond forces 
one to switch to a combination of analytic and numerical approaches to calculate virial
coefficients beyond second order. In cases of great interest such as spin-$1/2$ fermions,
one has to further break up the problem into the subspaces of fixed particle number; for example,
calculating $b_4$ requires solving the problem of $3+1$ particles (i.e. 3 spin up and 1 spin down,
and viceversa) as well as $2+2$ particles. The number of such subspaces naturally
proliferates with higher total particle number, and furthermore each subspace may present its
own difficulties.

To calculate $\Delta b_3$ there is only one distinct subspace that matters, 
namely $2+1$ particles (assuming that both particles have the same mass) and 
impressive exact analytic progress was made 
by several authors, notably the work of Leyronas~\cite{Leyronas2011PRA},
Kaplan and Sun~\cite{Kaplan2011PRL}, and Castin and colleagues~\cite{PhysRevA.85.033634, Endo2016JoPAMaT, Endo2016TEPJD, Gao2015EEL}, as well as the large effective range
expansion of Ngampruetikorn et al~\cite{PhysRevA.91.013606} (see also
the early work of Ref.~\cite{PhysRevB.67.174513} focusing on the unitary limit).

Leyronas organizes the calculation of $\Delta b_k$ around the VE of the density
equation of state, itself expressed as an integral over all momenta of the equal-time single-particle
Green's function in momentum space (i.e. the momentum distribution). Diagrams 
of various types are then identified at each order in $z$ (up to order $z^3$), where
contributions from the 2- and 3-body T matrix appear (the latter describing the atom-dimer
scattering). The resulting time integrals are converted into energy integrals, which are then 
evaluated analytically where possible and otherwise numerically. The resulting approach is
thus for the most part analytic and in principle exact and is in remarkable agreement
in the unitary limit with prior purely numerical results for $b_3$~\cite{PhysRevLett.102.160401}.

Other diagrammatic approaches also made interesting contributions. 
The work of Kaplan and Sun~\cite{Kaplan2011PRL}, which preceded Leyronas, 
starts from the density equation of state written as a momentum integral over the single-particle
Green's function (as Leyronas does), but rather than carrying out the Matsubara sum 
from the outset, it uses a Poisson summation to express the propagator directly as a power series in $z$. The
latter is then interpreted as a sum over winding number of worldlines around the compact 
imaginary time direction. The diagrams associated with each term in that expansion 
are referred to as `chronographs'. Adding the contributions from such chronographs
and accounting for systematic effects by extrapolation, very good agreement with 
prior numerical results~\cite{PhysRevLett.102.160401} for $b_3$ was obtained in the unitary limit.

Ngampruetikorn et al~\cite{PhysRevA.91.013606}, 
used an expansion around large effective range $R^*$ (compared to the thermal wavelength $\lambda_T$),
which allows them to examine up to the four-particle subspace diagrammatically, thus obtaining
numerical estimates for up to $\Delta b_4$. They focused on the unitary Fermi gas by interpolating between $R^* \gg \lambda_T$ and $\lambda_T /|a| \gg 1$, where $a$ is the
scattering length, and applied their method to the pressure, density, entropy, and spectral functions. Their interpolation results for $\Delta b_3$ and $\Delta b_4$ at unitarity agree with 
those obtained by other groups, including those presented here.
In Ref.~\cite{ParishEtAl}, Ngampruetikorn et al also studied the pairing correlations of the 2D Fermi gas up to third order in the virial expansion, additionally obtaining Tan's contact (we return to the expansion of this and other quantities below).

In Ref.~\cite{PhysRevLett.97.150401} Werner and Castin analyzed the (2+1)-body problem
of harmonically trapped spin-$1/2$ fermions at unitarity, obtaining their exact spectrum 
and eigenstates. Generalizing that work to the problem of 3+1 and 2+2 particles, Endo and Castin~\cite{Endo2016JoPAMaT, Endo2016TEPJD} (see also~\cite{Gao2015EEL})
calculated the value of $\Delta b_4$ as a function of the trapping frequency $\beta \omega$.
As we will show in Sec. 5.1, our non-perturbative determination turned out to be in remarkable agreement with their result, which they considered to be only a conjecture.

On the numerical side, some of the early works used exact diagonalization in hyperspherical coordinates~\cite{PhysRevA.85.033634, PhysRevLett.102.160401}, whereby a large number of 
eigenstates can be calculated and their energies summed over to calculate canonical partition 
functions, thus providing access to $b_3$ (and to some extent $b_4$, and with low accuracy $b_5$) for systems of cold atoms with short-range interactions.

In an outstanding numerical feat, Yan and Blume~\cite{PhysRevA.91.043607, PhysRevLett.116.230401} designed an ad hoc
Monte Carlo method to tackle the calculation of $\Delta b_4$ for fermions at unitarity, 
resulting in the first determination of this quantity with stochastic methods. 
Their calculation featured a harmonic trapping potential, which induces a temperature
dependence in $\Delta b_4$, which is expected to be temperature-independent in the unitary limit. (Below we will show a comparison between Yan and Blume's results and ours, when our method is generalized to include a harmonic trap.) The only important drawback of this work was the large uncertainty in the final result, induced by the increased stochastic noise as the trapping potential is removed. We will return to a discussion of this result below.

In spite of all of the above remarkable progress, it is evident that, due to the complexity of the 
$n$-particle quantum mechanical problem, a different kind of approach is needed if
one is to determine high-order virial coefficients with well controlled systematic error.
In particular, stochastic methods tend to have too large uncertainties to yield the accurate estimates needed to implement resummation techniques (we return to these below). On the
other hand, direct numerical methods such as exact diagonalization can be very powerful in providing detailed information (furnishing not only energies but also the associated eigenstates), but have not yet succeeded in accurately determining virial coefficients beyond third order.  
One of the main objectives of this paper is to present our work in developing and
applying a non-perturbative, semi-analytic, computational approach that is free of stochastic effects, beginning in the next section.
\\

\section{Homogeneous Fermi gases with a zero-range interaction}
\label{sec:base-case}

In this section we explain our method in detail and review its application to the 
simplest case, namely that of Fermi gases in homogeneous space, 
focusing on the VE for the pressure.

\subsection{Factorizing the transfer matrix}

The cornerstone of the approach is the Suzuki-Trotter factorization of the transfer matrix 
(i.e. the quantum version of the Boltzmann weight). To that end, the Hamiltonian is split into kinetic and potential energy terms, i.e.
\beq
\hat H = \hat T + \hat V,
\eeq
such that the simplest symmetric Suzuki-Trotter factorization is
\beq
e^{-\tau \hat H} = e^{-\tau \hat T/2} e^{-\tau \hat V} e^{-\tau \hat T/2} + O(\tau^3),
\eeq
where $\tau$ is in principle an arbitrary parameter, but we will define it
such that for some integer $N_\tau$, one has $\beta = \tau N_\tau$, i.e.
$\tau$ defines the imaginary time discretization. Note that, since our interest is in 
taking the trace of powers of $e^{-\tau \hat H}$, the same accuracy is
obtained for the symmetric decomposition as for its asymmetric counterpart
\beq
e^{-\tau \hat H} = e^{-\tau \hat T} e^{-\tau \hat V} + O(\tau^2).
\eeq
(Note: When calculating expectation values of operators, not mere traces, 
using the symmetric decomposition does make a difference.)
The above factorization step is always needed in our method, regardless of whether
the target system is in homogeneous space or in a trapping potential;
in this section we focus on the former, returning to the latter in a later section.

Using the above factorization, the objective is to calculate $Q_N$ for the desired 
particle content at progressively larger values of $N_\tau$.
The resulting $Q_N$ are then used to calculate the $b_k$, and the limit of large $N_\tau$
is taken at the end by extrapolation. The latter is an extrapolation to the continuous
imaginary-time limit.

The fundamental building blocks in the calculation of $Q_N$ are the matrix
elements of the factorized transfer matrix. 
For homogeneous non-relativistic spin-1/2 fermions with a zero-range interaction
one has
\beq
\hat T = \sum_{\sigma=\uparrow,\downarrow} \sum_{\mathbf{p}} \frac{\mathbf{p}_{}^2}{2m} \hat{n}_{\sigma}(\mathbf{p}) ,
\eeq
where $\hat{n}_{\sigma}(\mathbf{p})$ is the number density operator for particles of spin $\sigma$
and momentum $\bf p$, and we use \( m \equiv 1 \) for simplicity; moreover,
\beq
\hat V = g \ell^3 \sum_{\bf r} \hat{n}_{\uparrow}(\mathbf{r}) \hat{n}_{\downarrow}(\mathbf{r}), 
\eeq
where $\hat{n}_{\sigma}(\mathbf{r})$ is the number density operator for particles of spin $\sigma$
at position $\bf r$, and we have regularized the problem by putting it on a spatial lattice of spacing $\ell$. In practice, we renormalize this interaction by tuning it to reproduce
the exact value of $b_2$, as set by the Beth-Uhlenbeck formula reviewed above.

To calculate $Q_2$, for example, the desired matrix elements are given by
\begin{equation}
  \label{eq:6}
  \begin{aligned}
    \mathcal M_{11} & = \mel{\mathbf{p}_1 \mathbf{p}_2}{e^{- \tau \hat{T}} e^{- \tau \hat{V}}}{\mathbf{q}_1 \mathbf{q}_2} \\
    & = K(\mathbf{p}_1) K(\mathbf{p}_2)
    \left[ \delta_{{\bf q}_1 , {\bf p}_1} \delta_{{\bf q}_2 , {\bf p}_2} + 
C\  \delta_{{\bf q}_1 + {\bf q}_2 ,  {\bf p}_1 + {\bf p}_2 } \right], \\    
  \end{aligned}
\end{equation}
for the $(1+1)$-particle problem, where \( K({\bf p}) \) is the non-interacting, factorized Boltzmann weight
\begin{equation}
  \label{eq:7}
  K({\bf p}) = e^{- \tau {\bf p}^2 / (2m)},
\end{equation}
\( C \) is the coupling strength
\begin{equation}
  \label{eq:8}
  C = e^{\tau g / \ell^3 } - 1,
\end{equation}
and we use the notation
\begin{equation}
  \label{eq:9}
  \ket{\mathbf{P}} = \ket{\mathbf{p}_1 \mathbf{p}_2 \cdots \mathbf{p}_a \mathbf{p}_{a+1} \cdots \mathbf{p}_{a+b}},
\end{equation}
for the \( (a+b) \)-particle system, where \( \mathbf{p}_1 \) to
\( \mathbf{p}_a \) are for spin-\( \uparrow \) particles and
\( \mathbf{p}_{a+1} \) to \( \mathbf{p}_{a+b} \) for spin-\( \downarrow \) particles.

Similarly, for the $(2+1)$-particle problem, which enters in calculating $Q_3$, one obtains
\begin{equation}
  \label{eq:10}
  \begin{aligned}
     \mathcal M_{21} & = \mel{\mathbf{p}_1 \mathbf{p}_2 \mathbf{p}_3}{e^{- \tau \hat{T}} e^{- \tau \hat{V}}}{\mathbf{q}_1 \mathbf{q}_2 \mathbf{q}_3}\\
    & = K(\mathbf{p}_1) K(\mathbf{p}_2) K(\mathbf{p}_3)
    \left[ \delta_{{\bf q}_1 , {\bf p}_1} \delta_{{\bf q}_2 , {\bf p}_2} \delta_{{\bf q}_3 , {\bf p}_3} 
     + C \left( \delta_{{\bf q}_1 + {\bf q}_3 ,  {\bf p}_1 + {\bf p}_3 } + \delta_{{\bf q}_2 + {\bf q}_3 ,  {\bf p}_2 + {\bf p}_3 } \right)
    \right].    
  \end{aligned}
\end{equation}
Naturally, the complexity of the above matrix elements results mainly from the interaction
elements \( \mel{\mathbf{P}}{e^{- \tau \hat{V}}}{\mathbf{Q}} \) and rises rapidly with
the particle number. 

Another example is the \( (2+2) \)-particle problem, which pertains to $Q_4$, for which we obtain
\begin{equation}
  \label{eq:11}
  \begin{aligned}
     \mathcal M_{22} & = \mel{\mathbf{p}_1 \mathbf{p}_2 \mathbf{p}_3 \mathbf{p}_4}{e^{- \tau \hat{T}} e^{- \tau \hat{V}}}{\mathbf{q}_1 \mathbf{q}_2 \mathbf{q}_3 \mathbf{q}_4}\\
    & = K(\mathbf{p}_1) K(\mathbf{p}_2) K(\mathbf{p}_3) K(\mathbf{p}_4) \\
    & \qquad \times \left[ \delta_{{\bf q}_1 , {\bf p}_1} \delta_{{\bf q}_2 , {\bf p}_2} \delta_{{\bf q}_3 , {\bf p}_3} \delta_{{\bf q}_4 , {\bf p}_4} \right. \\
    & \qquad \quad + C\left(  \delta_{{\bf q}_1 + {\bf q}_3 ,  {\bf p}_1 + {\bf p}_3 }
    + \delta_{{\bf q}_2 + {\bf q}_3 ,  {\bf p}_2 + {\bf p}_3 }
    + \delta_{{\bf q}_1 + {\bf q}_4 ,  {\bf p}_1 + {\bf p}_4 }
    + \delta_{{\bf q}_2 + {\bf q}_4 ,  {\bf p}_2 + {\bf p}_4 } \right) \\
    & \left. \qquad \quad +\ C^2\left( \delta_{{\bf q}_1 + {\bf q}_3 ,  {\bf p}_1 + {\bf p}_3 } \delta_{{\bf q}_2 + {\bf q}_4 ,  {\bf p}_2 + {\bf p}_4 }
    + \delta_{{\bf q}_1 + {\bf q}_4 ,  {\bf p}_1 + {\bf p}_4 } \delta_{{\bf q}_2 + {\bf q}_3 ,  {\bf p}_2 + {\bf p}_3 } \right)
    \right].
  \end{aligned}
\end{equation}
%

\subsection{From transfer matrices to canonical partition functions}

In the above expressions for $\mathcal M_{21}$ and $\mathcal M_{22}$ we have
not implemented any symmetrization or antisymmetrization, which is 
needed to account for quantum statistics. Without approximation or loss of generality, that
operation can be carried out at the end of the calculation, i.e. upon taking the $N_\tau$-th power 
of the desired transfer matrix $\mathcal M_{ab}$, because the operators involved preserve the 
particle statistics. For example, in the fermionic case, the antisymmetrization yields
\beq
\label{eq:12}
Q^\text{F}_{21} = \tr^{}_\text{F}\left[ \mathcal M_{21}^{N_\tau} \right] = \frac{1}{2!} \sum_{a b c} \left\{ \left[ \mathcal M_{21}^{N_\tau} \right]_{abc, abc} - \left[ \mathcal M_{21}^{N_\tau} \right]_{abc, bac}\right\},
\eeq
whereas in the bosonic case one would have a symmetric form, namely
\beq
\label{eq:13}
Q^\text{B}_{21} = \tr^{}_\text{B}\left[ \mathcal M_{21}^{N_\tau} \right]
= \frac{1}{2!} \sum_{a b c} \left\{ \left[ \mathcal M_{21}^{N_\tau} \right]_{abc, abc} + \left[ \mathcal M_{21}^{N_\tau} \right]_{abc, bac}\right\}.
\eeq

To take the $N_\tau$-th power as shown above, we use automated tensor algebra (further details on this automation are presented below). Although 
the resulting number of terms is very large (on the order of $10^9$ for the cases we have explored),
it is manageable. Furthermore, each term is given by a multidimensional Gaussian integral (once the continuum and large-volume limits are taken) with an easily identifiable quadratic form. Since those integrals are 
easily evaluated using determinants, the entire process of algebraic manipulation and evaluation can be farmed out to massively distributed computing architectures as a large set of independent processes. In practice, we have been able to explore subspaces with up to 9 particles with several days of calculations on the Open Science Grid~\cite{osg07,osg09}. The number of particles one can analyze is of course limited by $N_\tau$; for instance with $10^6$ CPU hours, it is possible to study up to $N_\tau = 23$ for 3 particles, but only up to $N_\tau = 4$ for 9 particles.

Combining the results for the $Q^{}_{ab}$, one obtains the desired $b_k$. It should be noted that, in practice, one does not use these quantities directly but rather the changes induced by the interaction as given by $\Delta Q^{}_{ab}$ and $\Delta b_k$.

The main advantage of this method is that it is not a stochastic approach; in fact, it is closer to an
analytic approach, as it amounts to an automated, direct evaluation of a lattice field theory calculation.
The automated algebra allows us to resolve the volume cancellations that plague the 
evaluation of virial coefficients, which stem from combining $Q_{ab}$ for varying $a$ and $b$. 
As the latter scale as $V^{a+b}$ plus sub-leading terms, whereas the $b_k$ are 
volume independent, resolving those cancellations is both crucial and very difficult to achieve
with stochastic approaches.

\subsection{Computational details of automated algebra}

In this section we present a more detailed technical discussion of our automated algebra 
method to capture the general idea represented in our code.
The ultimate goal of the method is to evaluate the canonical partition functions as
shown in Eq.~(\ref{eq:12}) and (\ref{eq:13}), which involves three steps:
\begin{enumerate}
\item {\bf Term generation}: Expand the product \( M_{mj}^{N_{\tau}} \)
symbolically, which will yield a large number of terms as \( N_{\tau} \)
is increased.
\item {\bf Delta crunch}: Contract indices to saturate all Kronecker deltas, thus simplifying 
each term into a product of Gaussian functions,
namely the propagator \( K({\bf p}) \), by integrating out a subset of 
variables. This is the most computationally expensive step.
\item {\bf Gaussian integration}: For each term, take the summation over
  the rest of the variables and take the continuum limit, ultimately
  turning each term into a multidimensional Gaussian integral whose results
  are analytically available as the well-known formula
\begin{equation}
  \label{eq:14}
  \int \mathcal{D} \vec{x} \exp(- \frac{1}{2} \vec{x}^T A \vec{x} ) = \sqrt{\frac{(2 \pi)^n}{ \det A}},
\end{equation}
where \( n \) is the dimension of vector \( \vec{x} \).

\end{enumerate}

We now proceed to elaborate on the above steps.
\\

{\bf Step 1}\\
\indent As shown for instance in Eq.~\eqref{eq:11}, the kinetic energy appears
in our transfer matrix expressions as a prefactor, fully factorized across 
particles. The complexity of the problem lies in the interaction operators, of course.
In this first step, we expand the product that results from such operators when
$N_\tau$ factors are present: 
\begin{equation}
  \label{eq:15}
    \prod_{i=1}^{N_{\tau}} \mel{\mathbf{P}^{(i)}}{e^{-\tau \hat{V}}}{\mathbf{P}^{(i+1)}}
    = \prod_{i=1}^{N_{\tau}} [1 + C f_1(\mathbf{P}^{(i)}, \mathbf{P}^{(i+1)}) + C^2
    f_2(\mathbf{P}^{(i)}, \mathbf{P}^{(i+1)}) + \cdots ]
\end{equation}
where
\beq
\mathbf{P}^{(i)} = \{ \mathbf{p}^{(i)}_{1}, \mathbf{p}^{(i)}_{2}, \cdots
\mathbf{p}^{(i)}_{M}, \mathbf{p}^{(i)}_{M+1}, \cdots \mathbf{p}^{(i)}_{M+N} \},
\eeq
is the \( i \)-th complete set inserted, \( f_i(\cdot) \) is a function
containing Kronecker \( \delta \)'s that implement all possible interactions
(that result from the original two-body force) along imaginary time, and
the ellipses in Eq.~(\ref{eq:15}) indicate a polynomial of degree at most
\( \min(M, N) \), where \( M \) and \( N \) are the numbers of spin-up and
spin-down particles respectively.
The result of this step is a high-degree polynomial in the coupling strength \( C \).

To evaluate the matrix sums with different subscripts [e.g. as in Eq.~(\ref{eq:12})], 
thus implementing the pertinent quantum statistics,
we implement different boundary conditions on \( \mathbf{P}^{(1)} \) and
\( \mathbf{P}^{(N_{\tau})} \). For example, in Eq.~(\ref{eq:12}), the first
term on the right-hand side is a normal trace, which we obtain by imposing 
a periodic boundary condition
\( \mathbf{p}^{(1)}_i = \mathbf{p}^{(N_{\tau})}_i, \forall i = 1, 2, 3 \).
On the other hand, the second term
is a kind of ``shifted'' trace, obtained by setting the boundary condition 
\( \mathbf{p}^{(1)}_1 = \mathbf{p}^{(N_{\tau})}_2, \mathbf{p}^{(1)}_2 = \mathbf{p}^{(N_{\tau})}_1 \) and \( \mathbf{p}^{(1)}_3 = \mathbf{p}^{(N_{\tau})}_3 \).

We have considered so far the complete expansion of the product. However,
thanks to the cyclic property of the trace, only a subset
of the complete expansion needs to be evaluated. 
For example, for \( N_{\tau} = 2 \), there are two terms
\( f_1(\mathbf{P}^{(1)}, \mathbf{P}^{(2)}) f_2(\mathbf{P}^{(2)},
\mathbf{P}^{(3)}) \) and
\( f_2(\mathbf{P}^{(1)}, \mathbf{P}^{(2)}) f_1(\mathbf{P}^{(2)},
\mathbf{P}^{(3)}) \) at \( \mathcal{O}(C^3) \), 
which are equivalent under the cyclic variable substitution 
\( \mathbf{P}^{(1)} \to \mathbf{P}^{(2)} \),
\( \mathbf{P}^{(2)} \to \mathbf{P}^{(3)} \),
\( \mathbf{P}^{(3)} \to \mathbf{P}^{(1)} \).
Such a property defines a mathematical object called ``combinatorial
necklace'' and the goal of this first step is then to generate all possible
necklaces of functions \( f_i \), for which we used the algorithm developed
in Ref.~\cite{ruskey1992generating}.
\\

{\bf Step 2}
\\
\indent 
Once we have expanded the product of interaction operators, each term is in the
form of a product of propagators and \( \delta \)'s as
\begin{equation}
  \label{eq:16}
  K(\mathbf{P}^{(1)}) K(\mathbf{P}^{(2)}) K(\mathbf{P}^{(3)}) \cdots  \times \Delta(\mathbf{P}^{(1)}, \mathbf{P}^{(2)}, \mathbf{P}^{(3)}, \cdots )
\end{equation}
where \( K(\mathbf{P}^{(i)}) \) is a shorthand for the kinetic-energy product
\( K(\mathbf{p}^{(i)}_1) K(\mathbf{p}^{(i)}_2) \cdots K(\mathbf{p}^{(i)}_{M+N})
\), and
\( \Delta(\mathbf{P}^{(1)}, \mathbf{P}^{(2)}, \mathbf{P}^{(3)}, \cdots ) \) is a
product of Kronecker \( \delta \)'s from the combination of
\( f_i(\mathbf{P}^{(i)}, \mathbf{P}^{(i+1)}) \) functions. 
Equations~(\ref{eq:6}), (\ref{eq:10}) or (\ref{eq:11}) exemplify what one would 
obtain in the simplest case of \( N_{\tau} = 1 \). Thus, the output of this
step is effectively the tensor contraction (for all internal indices; not the trace indices) 
of such \( N_{\tau} = 1 \) results for $N_\tau$ as large as needed.

The second step of the method is to carry out the sums on such a term
over all momentum variables from \( \mathbf{P}^{(2)} \) to
\( \mathbf{P}^{(N_{\tau} - 1)} \), i.e. all intermediate complete sets
inserted. This operation is called ``Delta crunch'' as it will reduce the
\( \Delta \) function by substituting all available momentum variables,
i.e. ``crunching'' the \( \delta \)'s into the \( K \) factors. To this
end, we loop through the \( \delta \)'s in the \( \Delta \) function one at a
time, and perform variable substitution in both \( K \) and
\( \Delta \). To provide efficiency in variable
substitutions, the \( K \) and the \( \delta \)'s are
represented as a hashmap, which offers a \( \mathcal{O}(1) \) time in lookup and
modification.
\\

{\bf Step 3}
\\
\indent Once the Delta crunch step is complete, the resulting expression
will be a product of kinetic energy factors with corresponding variable
substitutions, e.g.
\begin{equation}
  \label{eq:17}
  K(\mathbf{p}^{(1)}_1) K(\mathbf{p}^{(1)}_2) K(\mathbf{p}^{(1)}_3) K(\mathbf{p}^{(1)}_1 + \mathbf{p}^{(1)}_2 - \mathbf{p}^{(1)}_3) \cdots .
\end{equation}
Recalling that $K({\bf p})$ is a Gaussian function, the evaluation of the
summation over the remaining momentum variables is easiest carried out in the 
continuum limit. To that end, we convert the above product into a quadratic form
\begin{equation}
  \label{eq:18}
  \exp(- \frac{\tau}{2 m} \vec{p}^T \mathbf{A} \vec{p} )
\end{equation}
where \( \vec{p} \) contains all the momentum variables. The matrix
\( \mathbf{A} \) is symmetric and positive-definite, such that one can use 
Cholesky decomposition to evaluate the determinant, which is computationally more
efficient and numerically more stable than LU decomposition.
\\


{\bf Parallelization}

Before concluding this section, we want to add one more technical note on
the parallelization of our method. Compared to more conventional methods
like QMC, ours is much easier to parallelize. All the terms generated in
the first step are independent from each other, which means they can be
evaluated in fully parallel fashion with little or no communication
overhead among processes. Moreover, the evaluation of each term is cheap as
it does not involve complicated linear algebra operations, and so it is
suitable to run on any number of CPU cores. These features make our method
ideal to run on a distributed, heterogeneous computing cluster, such as the
Open Science Grid or the Folding@home project, where the computational
power is unevenly distributed across nodes, in contrast with traditional
supercomputers, while the number of available cores can be much higher.
\\

\subsection{Selected results}

Using the method presented above, which resulted from a sequence of
prior studies~\cite{PhysRevA.98.053615, PhysRevA.100.063626, Hou2019PRA, Czejdo2020PRA}, we tackled the problem of calculating, 
with as high precision as possible, the virial coefficients up to $\Delta b_5$ for 
homogeneous spin-$1/2$ Fermi gases with an attractive contact interaction, for 
varying coupling strengths. For illustrative purposes, we focus
here on the 3D case~\cite{Hou2020PRL}, but extensions
to other dimensions were also studied~\cite{Hou2020PRA}.

\begin{figure}[h]
  \centering
  \includegraphics[width=\linewidth]{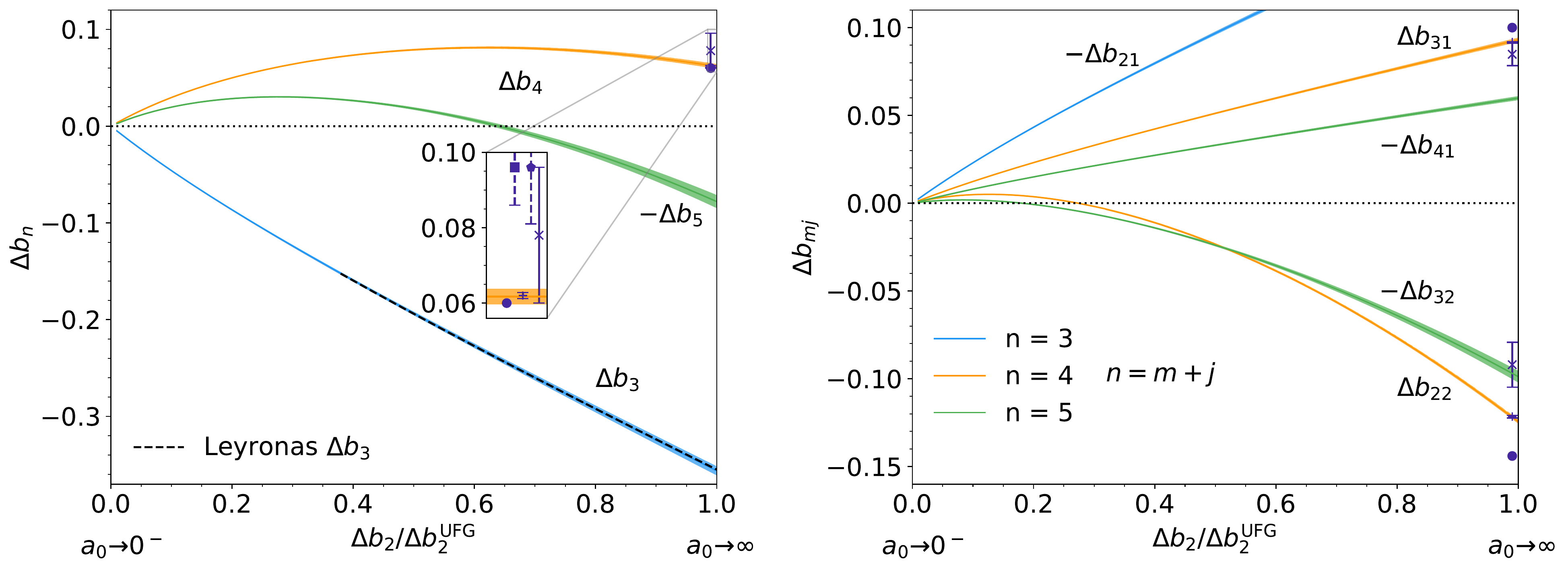}    
  \caption{
  \label{fig:homo-dbn}
    {\bf Left}: $\Delta b_n$ for $n=3, 4, 5$, from weak coupling (scattering length $a_0 \to 0$)
  to unitarity ($a_0 \to \infty$) as parameterized by $\Delta b_2/\Delta b_2^\text{UFG}$, where
  $\Delta b_2^\text{UFG} = 1/\sqrt{2}$ is the value of $\Delta b_2$ at unitarity. The inset shows
  a zoom into the region around unitarity, where several experimental and theoretical estimates are shown for comparison (see main text and Ref.~\cite{Hou2020PRL} for details).
  {\bf Right}: Subspace contributions \( \Delta b_{ij} \).
  }
\end{figure}

In Fig.~\ref{fig:homo-dbn}, we present the coefficients \( \Delta b_k \) for
\( k = 3, 4, 5\) (left panel) and the corresponding subspace contribution \( \Delta b_{ij} \) 
(right panel). The inset on the left panel shows a comparison with 
experiments~\cite{Salomon, MITExpKu2012S}
and theory~\cite{PhysRevA.91.013606, Endo2016JoPAMaT, PhysRevLett.116.230401}. 

Besides the excellent agreement with the $\Delta b_3$ results of
Ref.~\cite{Leyronas2011PRA}, the above provide a precise non-perturbative,
non-stochastic determination of $\Delta b_4$ and $\Delta b_5$. The results show that,
due to the competing subspace contributions (right panel), both
$\Delta b_4$ and $\Delta b_5$ are non-monotonic as a function of the coupling
strength. In addition, when approaching the unitary limit it is evident
that $\Delta b_5$ becomes comparable in magnitude to $\Delta b_4$. It is this
property that complicates the experimental determination of $\Delta b_4$ at
strong coupling (dashed error bars in inset of left panel), as one is
forced to assume that $\Delta b_5$ is comparatively small, which is not the
case.  The agreement for $\Delta b_4$ with the estimates of
Refs~\cite{PhysRevA.91.013606, Endo2016JoPAMaT, PhysRevLett.116.230401},
which use entirely different methods (among them and with the present
work), further supports the accuracy and precision of the method. Note,
however, the wide disparities in results for $\Delta b_{22}$ compared to
$\Delta b_{31}$, which indicate that the $(2+2)$-particle subspace is a
considerably more difficult problem to tackle than the polarized $(3+1)$
subspace.

\section{Generalization to other systems}
\label{sec:gener-other-syst-1}

\subsection{Harmonic traps}
\label{sec:harm-trapp-syst}

In the case of harmonically trapped particles, it is convenient to split the Hamiltonian
into a noninteracting harmonic oscillator piece $\hat H_0$, and the interaction $\hat V$
in order to perform the Suzuki-Trotter factorization, such that
\beq
\hat H = \hat H_0 + \hat V,
\eeq
where $\hat H_0 = \hat T + \hat V_\text{HO}$ and 
\beq
\hat V^{}_\text{HO} = \int\! d^3r\, \frac{1}{2} m \omega^2 {\bf r}^2  \left [\hat n_{\uparrow}^{}({\bf r}) + \hat n_{\downarrow}^{}({\bf r}) \right],
\eeq
is the harmonic trapping potential, and therefore
\beq
e^{-\tau \hat H} = e^{-\tau \hat H_0/2} e^{-\tau \hat V} e^{-\tau \hat H_0/2} + O(\tau^3).
\eeq
The main advantage of this choice of splitting of $\hat H$ is that, as we will see below,
the resulting factorization of the transfer matrices can be easily written in terms of the so-called Mehler kernel~\cite{PhysRevA.100.063626}.
It should be pointed out that the other possible factorization where one combines $\hat V_\text{HO}$
with $\hat V$ rather than with $\hat T$ is an interesting possibility that one could explore
entirely in momentum space (as in the homogeneous cases of the previous section). 
However, such an approach effectively results in a more complicated interaction made out of a one-body piece and a two-body 
piece with independent coupling constants, which must be kept track of throughout the calculation,
with the concomitant increase in computational complexity. We will return to a similar situation
below, namely the case of attractive Bose gases, which require two- and three-body forces
for stability reasons.

As anticipated, the transfer matrices can be written conveniently in terms of the Mehler kernel.
For example, when examining the $(1+1)$-particle subspace, we obtain 
\bea
\mathcal M_{11} &=& \langle {\bf x}_1,{\bf x}_2 |  e^{- \tau (\hat{T} + \hat V^{}_\text{HO})} e^{-\tau \hat{V}} | {\bf y}_1,{\bf y}_2 \rangle \nonumber \\ 
&=&
\rho({\bf x}_1,{\bf y}_1)\rho({\bf x}_2,{\bf y}_2) \left[ \mathbbm{1} + C \delta({\bf y}_1-{\bf y}_2)\right],
\eea
where $\rho({\bf x},{\bf y})$ is the Mehler kernel
\beq
\label{Eq:MehlerKernel}
\rho({\bf x},{\bf y}) = \frac{1}{\lambda^3_T}\left[\frac{\beta \omega}{\sinh(\tau \omega)}\right]^{3/2}\exp[-{\bf Z}^T B {\bf Z}],
\eeq
which encodes the coordinate representation of the transfer matrix of a single-particle in the harmonic potential $\hat V^{}_\text{HO}$.
Here, ${\bf Z}^T = ({\bf x}^T/\lambda_T, {\bf y}^T/\lambda_T)$, and
\bea
B = 
\frac{\pi \beta \omega}{\sinh(\tau \omega)}
\left(
\begin{array}{cc}
\cosh(\tau \omega)\mathbbm{1} & -\mathbbm{1}\\
-\mathbbm{1}  & \cosh(\tau \omega)\mathbbm{1}
\end{array}
\right),
\eea
where $\mathbbm{1}$ is a $3\times3$ unit matrix. 
Similarly, for the $(2+1)$-particle subspace, we find
\bea
\mathcal M_{21} &=& \rho({\bf x}_1,{\bf y}_1)\rho({\bf x}_2,{\bf y}_2)\rho({\bf x}_3,{\bf y}_3) 
\left\{ \mathbbm{1} + C [\delta({\bf y}_1-{\bf y}_3)  + \delta({\bf y}_2-{\bf y}_3)]\right\}.
\eea
Note that, as in a previous section, we do not include any symmetrization or
antisymmetrization at this stage, as quantum statistics can be accounted for after
the $N_\tau$-th power is taken.
Further transfer matrices for up to 5 particles can be found in Ref.~\cite{Hou2021PRRFourth}.

As in the homogeneous case, we rely on known, analytic results for $\Delta b_2$ in
order to renormalize the interaction.
In a harmonic trapping potential \( \Delta b^T_2 \) is given at unitarity by \cite{Liu2013PR}
\begin{equation}
  \label{eq:19}
  \Delta b^T_2 = \frac{1}{4} \sech(\frac{\beta \omega}{ 2}).
\end{equation}
We note that, for arbitrary order, the trapped virial coefficients \( \Delta b_n^T \) in
the high-temperature (or low-frequency) limit \( \beta \omega \to 0 \) is connected with their
homogeneous counterparts via
\begin{equation}
  \label{eq:20}
  \Delta b_k = k^{3/2} \Delta b^T_k(\beta \omega \to 0),
\end{equation}
which is useful to know when comparing homogeneous results with
low-frequency extrapolations from the trapped case.

Using our framework at leading order (\( N_{\tau} = 1 \)) in $d$ spatial dimensions, it is
not difficult to obtain analytic formulas such as
\beq
\Delta b_{2} = \Delta b_{11} =  \frac{1}{2} \frac{C}{\lambda_T^d}\left[\frac{\beta \omega}{2 \sinh(\beta \omega)} \right]^{d/2},
\eeq
\beq
\Delta b_{21} =  -\frac{\Delta b_{2}}{\left [2\cosh(\beta \omega)+1 \right ]^{d/2}},
\eeq
\beq
\Delta b_{31} = \frac{2^{-d/2}\Delta b_{2}}{ \cosh^{d/2}(\beta \omega) \left [2\cosh(\beta \omega) + 1\right]^{d/2}},
\eeq
and
\bea
\Delta b_{22} &=& \frac{2^{-3d/2} \Delta b_{2}}{\cosh^{d/2}(\beta \omega) \cosh^{d}(\beta \omega/2)} \times\\
&&\!\!\!\!\!\!\!\!  \left\{ 1 + 2^{d/2} \Delta b_{2} \left[\cosh^{d/2}(\beta \omega) - 2^{d/2+1} \cosh^{d}(\beta \omega/2)\right] \right\}
\nonumber.
\eea
Following the same approach outlined in previous sections, we worked out the next-to-leading order (\( N_{\tau} = 2 \)) formulas, which the reader can find in Ref.~\cite{Hou2021PRRFourth}.
With automated algebra, these can be pushed to much higher $N_\tau$ to obtain
estimates for $\Delta b_k$ which are then extrapolated to large $N_\tau$.

\begin{figure}[h]
  \centering
  \includegraphics[width=\linewidth]{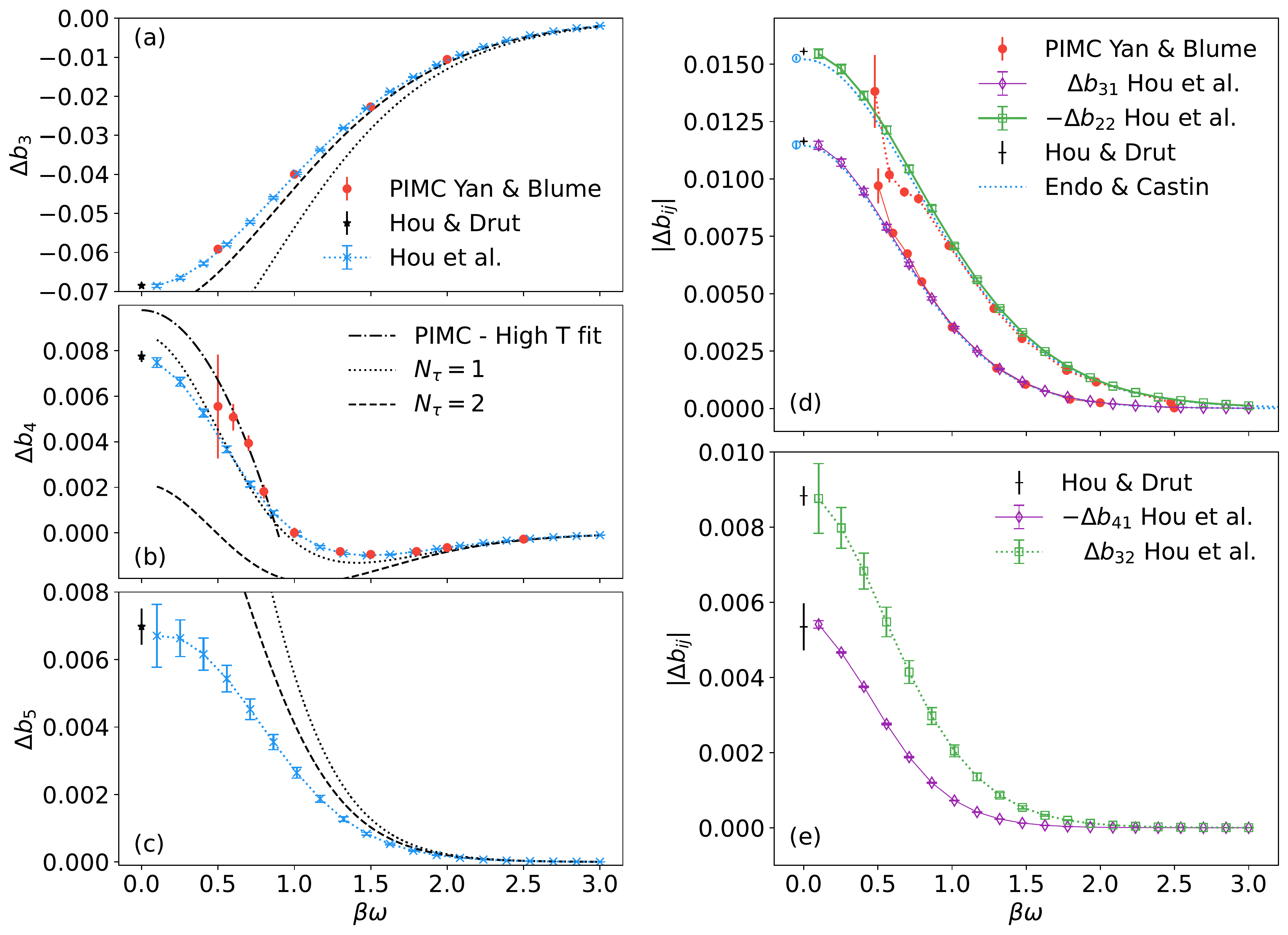}    
  \caption{{\bf(a) to (c)}: \( \Delta b_3 \), \( \Delta b_4 \) and
    \( \Delta b_5 \) as function of \( \beta \omega \) respectively. Our calculations are
    presented as blue crosses with error bars. The dotted and dashed black
    lines are the leading- and next-leading-order results. The black stars
    at \( \beta \omega \to 0 \) is the results of the homogeneous system. The red
    solid circles in panel (a) and (b) show the results by Yan and Blume
    \cite{PhysRevLett.116.230401}. In panel (b), the dash-dotted black line
    is the high-temperature fitting from the same work. {\bf(d) and (e)}:
    Subspace contribution \( \Delta b_{ij} \) for (d) \( n = 4 \) and (e)
    \( n = 5 \). The open green squares with dotted lines represent the
    \( \Delta b_{x1} \) contribution, and the open purple diamond for the
    \( \Delta b_{x2} \) contribution. The black bar are the results of the
    homogeneous system. In panel (d), The dotted red lines are the
    theoretical conjecture from Ref.~\cite{Endo2016JoPAMaT}, and the open
    circle is to emphasize the results in the \( \beta \omega \to 0 \) limit. The PIMC
    results from Ref.~\cite{PhysRevLett.116.230401} are shown as red
    circles.}
  \label{fig:sho-dbn}
\end{figure}

In Fig.~\ref{fig:sho-dbn}, we show our results at \( N_{\tau} = 1, 2 \) for the
full-space contribution \( \Delta b_k \), and those extrapolated to
\( N_{\tau} \to \infty \) limit for both \( \Delta b_k \) and the corresponding subspace
contribution \( \Delta b_{ij} \).
We find very good agreement between the extrapolated results and the
Monte Carlo calculations of Ref.~\cite{PhysRevLett.116.230401} for large trapping 
frequency \( \beta \omega \). As \( \beta \omega \) approaches \( 0 \), our
results are smoother compared to the Monte Carlo results, likely due to the volume
cancellations that we are able to resolve analytically but which induce noise in the Monte Carlo
method.
Although the leading- and next-to-leading-order results deviate from the
expected curve, they capture most qualitative features and may shed light
into higher-order virial coefficients where the computational costs are 
too large for a full calculation at large $N_\tau$.
Lastly, note that for \( \Delta b_4 \), the leading-order result is closer to
the extrapolated curve compared to the next-leading-order result. This
  indicates that the discretization error from the Suzuki-Trotter
  decomposition is not monotonic in \( N_{\tau} \). 
  In other words, as \( N_{\tau} \) increases, we may expect worse
results before the asymptotic regime is approached. The onset of that
regime is of course coupling dependent. Therefore, at a given interacting strength 
it is essential to investigate \( N_{\tau} \) as high as possible in order to obtain 
quantitatively correct results.

\subsection{Neutron Matter}
\label{sec:neutron-matter}

The outermost crust of a neutron star is a low-density regime for which
the unitary Fermi gas results presented above can be regarded as an approximation.
To go beyond the unitary limit, a realistic model for neutron matter must include 
interaction range effects. In this section we elaborate on a possible implementation of a
finite-range interaction in the context of our method. To that end, we anticipate that it
will be essential to expand the matrix elements of $e^{-\tau \hat V}$ as a sum of
Gaussians, as that would take advantage of the computational framework developed in 
the simpler cases discussed above.

As a reminder, for a contact interaction in the two-particle subspace one has 
\beq
\langle {\bf q}_1 {\bf q}_2 | e^{-\tau \hat V} | {\bf p}_1 {\bf p}_2 \rangle = 
\delta_{{\bf q}_1 , {\bf p}_1} \delta_{{\bf q}_2 , {\bf p}_2} + 
C\  \delta_{{\bf q}_1 + {\bf q}_2 ,  {\bf p}_1 + {\bf p}_2 },
\eeq
where $C$ includes the lattice spacings and the bare coupling. Naturally, the first term
represents a noninteracting piece and the second term encodes the interaction effects.
A contact interaction in coordinate space has constant matrix elements in momentum space,
but in the presence of a finite range there will be nontrivial momentum dependence and therefore 
we generalize the above via
\beq
C \to C({\bf q}_r, {\bf p}_r) = \sum_k C_k\ e^{- \tau \lambda_{1,k} ({\bf q}_r^2 + {\bf p}_r^2) + 
\tau \lambda_{2,k} {\bf q}_r \cdot {\bf p}_r},
\eeq
where ${\bf q}_r ={\bf q}_1 - {\bf q}_2$  and ${\bf p}_r = {\bf p}_1 - {\bf p}_2$. 
In the limit \( \lambda_{1},\lambda_{2} \to 0 \), we reproduce the zero-range model.
This type of expansion can be further generalized by including powers of ${\bf q}_r^2$,
${\bf p}_r^2$ and ${\bf q}_r \cdot {\bf p}_r$ as a prefactor rather than in the exponent,
which can be used to account for more features of the interaction. However, such a 
generalization should be used sparingly or not at all if possible, as it would dramatically
(specifically, factorially) increase the number of terms that result in the expansion, by 
virtue of Wick's theorem.

\subsection{Unitary Bose gas}
\label{sec:unitary-bose-gas}

The case of Bose gases with attractive interactions, in particular close to the universal regime
of the unitary limit, requires special care. Given the absence of Pauli exclusion, attractively interacting bosons undergo Thomas collapse~\cite{PhysRev.47.903}, i.e. they are unstable. 
Moreover, close to unitarity they display the Efimov effect~\cite{Efimov1, Efimov2}.
To properly renormalize such a system,
a repulsive three-body interaction is required~\cite{PhysRevLett.82.463}. The latter introduces a new dimensionful parameter
that is sensitive to the ultraviolet cutoff and must be fixed by using some known physical quantity.

To account for the above, we consider an interaction of the form
\beq
\hat V = -\frac{g_2}{2!} \sum_{\bf r} (a^\dagger_{\bf r})^2 (\hat a^{}_{\bf r})^2 - \frac{g_3}{3!} \sum_{\bf r} (\hat a^\dagger_{\bf r})^3 (\hat a^{}_{\bf r})^3,
\eeq
where we have used the normal-ordered form and $\hat a^{\dagger}_{\bf r}$, $\hat a^{}_{\bf r}$
are the bosonic creation and annihilation operators at point $\bf r$.

In previous sections we used $\Delta b_2$ to renormalize a two-body contact interaction.
It is therefore natural in the present case to use $\Delta b_2$ and $\Delta b_3$ for the same purpose.
Since $\Delta b_2$ involves only the two-particle subspace, we use it to renormalize the two-body 
interaction in the usual way before proceeding to $\Delta b_3$, which depends on both the two- and three-particle subspaces and thus determines the three-body coupling.
In the bosonic unitary limit, the three-body parameter induces a temperature dependence on 
$\Delta b_3$, which was calculated exactly in Ref.~\cite{Castin2013CJoP}.

Calculating $\Delta b_2$ using the above interaction in $d$ spatial dimensions and $N_\tau = 1$
we obtain
\beq
\Delta b_2 = \frac{\Delta Q_2}{Q_1}= \beta g_2 \frac{Q_1}{V},
\eeq
where $V = L^d$ is the $d$-dimensional volume, such that in the (spatial) continuum limit
\beq
\Delta b_2 \to \frac{1}{2\pi} \frac{g_2}{\lambda_T^{d-2}},
\eeq
where we used that in that limit ${Q_1}/{V} \to \lambda_T^{-d}$.

On the other hand, a calculation of $\Delta b_3$, also in $d$ spatial dimensions and $N_\tau = 1$, yields
\beq
\Delta b_3 = \frac{\Delta Q_3}{Q_1} - Q_1 \Delta b_2 =  \frac{\beta g_2}{2!} \; 4 \frac{Q_1(2\beta)}{V} + \frac{\beta g_3}{3!} \; 6 \frac{Q_1^2}{V^2},
\eeq
which in the (spatial) continuum limit becomes
\beq
\Delta b_3 \to \frac{1}{2^{d/2 - 1}}\frac{1}{2 \pi}\frac{g_2}{\lambda_T^{d-2}} + \frac{3}{\pi} \frac{g_3}{\lambda_T^{2d - 2}}.
\eeq
We thus see that, as anticipated, $\Delta b_2$ and $\Delta b_3$ determine $g_2$ and $g_3$.
At the $N_\tau = 1$ level of this calculation, the temperature dependence of $\Delta b_2$ and $\Delta b_3$ 
will typically induce a temperature dependence in $g_2$ and $g_3$. More explicitly, we may invert the above
results to obtain the dimensionless form
\bea
\beta g_2 \lambda_T^{-d}&=& \Delta b_2, \\
\beta g_3 \lambda_T^{-2d}&=& \Delta b_3 - \frac{1}{2^{d/2 - 1}}\Delta b_2.
\eea

Armed with these answers, the renormalization program is complete and we may proceed to calculate
and predict higher-order virial coefficients.
Our leading order ($N_\tau = 1$) for the fourth-order bosonic virial coefficient $\Delta b_4$ is
neatly expressed in terms of a relationship with $\Delta b_2$ and $\Delta b_3$:
\beq
\label{eq:21}
\Delta b_4 = \left (3^{-d/2}2  + 2^{-d} - \frac{3}{2^{d - 1}} \right )\Delta b_2 + \left (2^{-d/2 + 1} + 2^{-2d - 1} \right)\left (\Delta b_2\right )^2 + \frac{3}{2^{d/2}} \Delta b_3.
\eeq

\begin{figure}[h]
  \centering
  \includegraphics[width=.6\linewidth]{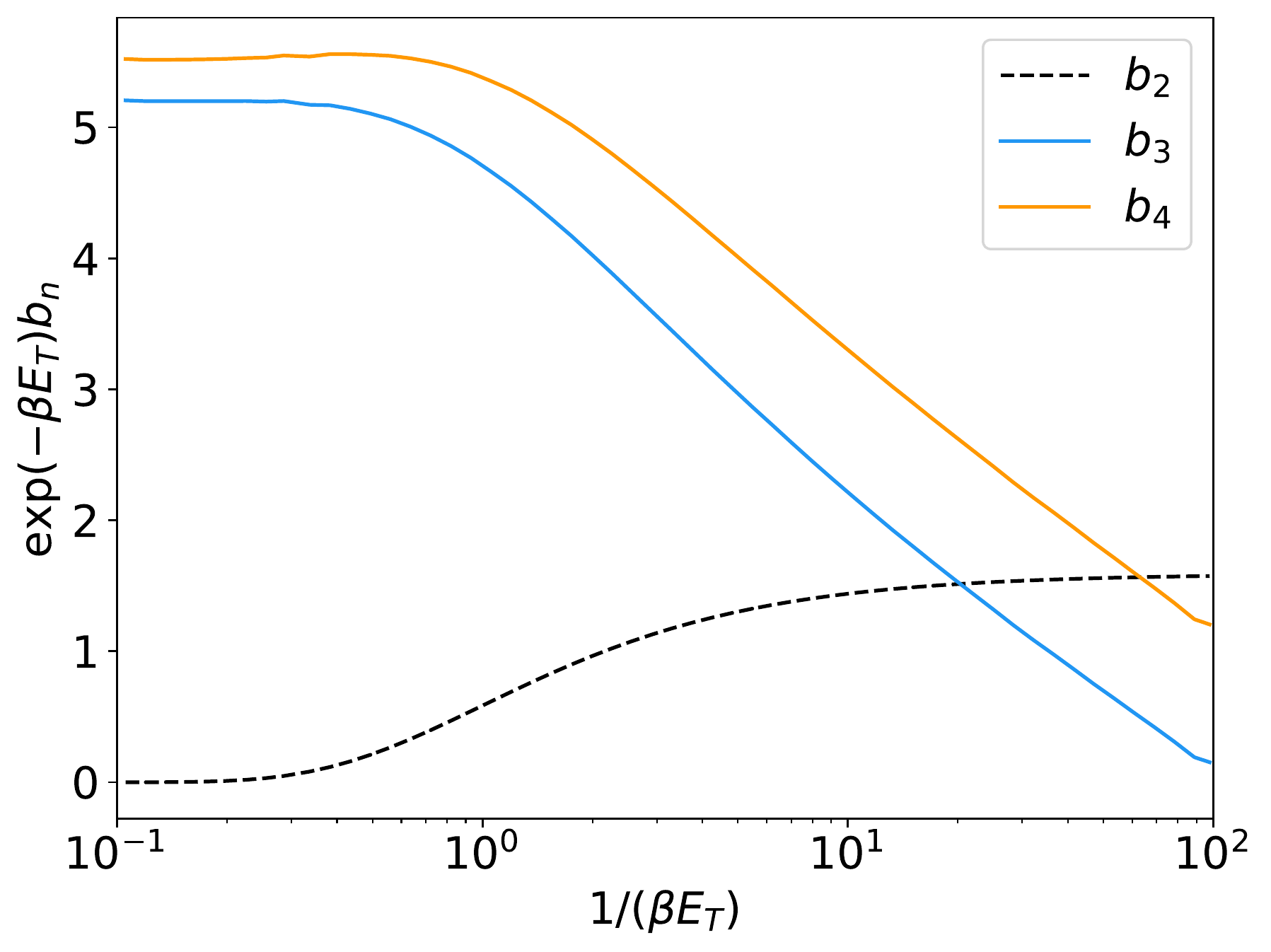}    
  \caption{Scaled virial coefficients \( \exp(-\beta E_T) b_n \) as a function
    of \( (\beta E_T)^{-1} \) where \( E_T \) is the ground state energy of the
    trimer. The black dashed line corresponds to the constant second order
    value $b_2 = (9\sqrt{2})/8$. The blue line is the scaled \( b_3 \) from
    Ref.~\cite{Castin2013CJoP}, which we used in Eq.~(\ref{eq:21}) to obtain the
    orange line showing \( b_4 \).}
  \label{fig:ubg}
\end{figure}

In Fig.~\ref{fig:ubg} we show the above result applied to $d=3$ at unitarity, where 
$\Delta b_2 = \sqrt{2}$ and $\Delta b_3$ is as determined in Ref.~\cite{Castin2013CJoP}.
Our $N_\tau = 1$ approximation is likely a rather crude one, but it should be asymptotically
correct for small $\beta E_T$.

\section{Pressure, density, and generalization to other observables}

So far, we limited our scope to the expansion for pressure, but the virial expansion is much more 
general and can be applied to a wide range of observables. In this section, we present the 
generalization of our approach to the Tan contact, the momentum distribution, and the structure factor.

\subsection{From pressure to density and Tan's contact}
\label{sec:PressureDensityContact}

By way of the grand thermodynamic potential, the \( b_n \) grant access to all 
thermodynamic quantities, at least in principle. One of
the most basic quantities besides the pressure is the density, which is given by
\beq
\label{eq:22}
n = \pdv{\ln \mathcal{Z}}{\ln z} = n_0 + \frac{2}{\lambda_T^d} \sum_{k=2}^{\infty} m \Delta b_k z^k,
\eeq
where \( n_0 \) is the density at absence of interaction, and similarly in
the presence of finite polarization (i.e. different fugacities $z_\uparrow$ and
$z_\downarrow$ for each spin),
\begin{equation}
  \label{eq:23}
  n_{s} = n_{s,0} + \frac{2}{\lambda_T^d} \sum_{k=2}^{\infty} \sum_{ \substack{i,j>0\\i+j=k}} [i \delta_{s,\uparrow} + j \delta_{s, \downarrow}] \Delta b_{ij} z^i_{\uparrow} z^j_{\downarrow},
\end{equation}
where \( s = \uparrow, \downarrow \) for each spin.

Beyond global thermodynamic quantities, the Tan contact can also be
investigated through the virial expansion. For systems with short-range two-body forces,
Tan's contact represents the probability of finding two particles at the same spatial location.
For that reason, it captures the short-distance behavior of all correlation functions.
In particular, the momentum distribution in such systems decays at large momentum $k$
as 
\beq
n \sim \mathcal I /k^4,
\eeq
where $\mathcal I$ is the contact.
Thus, one way to measure or calculate the contact is to determine the
large-momentum tail of the momentum distribution. In practice, however, it
has proven more efficient to use the so-called adiabatic relation, whereby
$\mathcal I$ is obtained from the variation of the grand thermodynamic potential
with the coupling strength. Below we
focus on the 3D case, but similar equations can be derived in one- and
two-spatial dimensions and readers are referred to Ref.~\cite{Hou2020PRA} for
more details. 

In three spatial dimensions, the VE for the contact is obtained via
\beq
  \label{eq:24}
  \mathcal{I}  = - \frac{4 \pi}{\beta} \pdv{(\beta \Omega)}{a_0^{-1}}
  = \frac{4 \pi}{\sqrt{\beta}} \pdv{\ln \mathcal{Z}}{\lambda}
  = \frac{4 \pi}{\beta} Q_1 \lambda_T \sum_{k=2}^{\infty} c_k z^k,
\eeq
where
\beq
\label{eq:25}
c_k = \frac{1}{\sqrt{2 \pi}} \frac{\partial \Delta b_k}{\partial \lambda},
\eeq
are the virial coefficients of $\mathcal{I}$.
To evaluate the partial derivative, the most straightforward method is to
apply the chain rule as
\begin{equation}
  \label{eq:26}
  \begin{aligned}
    \pdv{\Delta b_k}{\lambda} = \pdv{\Delta b_{k}}{\Delta b_2} \pdv{\Delta b_2}{\lambda},
  \end{aligned}
\end{equation}
where the first derivative can be calculated numerically and the second is
analytically given by the Beth-Uhlenbeck formula as
\begin{equation}
  \label{eq:27}
\pdv{\Delta b_2}{\lambda} = \sqrt{\frac{2}{\pi}} + \sqrt{2} \lambda e^{\lambda^2} [1 + \erf(\lambda)].
\end{equation}
Thanks to the analytical nature of our method, one can improve the
accuracy of the numerical derivative without repeating calculations.
Alternatively, one can take advantage of the analytic form of
\( \Delta b_k \) from our method, which is a polynomial in terms of coupling
strength \( C \)
\begin{equation}
  \label{eq:28}
  \Delta b_k = \sum_{l=1}^{l_{\mathrm{max}}} A_l C^{l},
\end{equation}
where \( l_{\mathrm{max}} = \min(M, N) \cdot N_{\tau} \) is the degree 
of the polynomial. The derivative can then be treated as a polynomial in \( C \) of
degree \( l_{\mathrm{max}} - 1 \) as
\begin{equation}
  \label{eq:29}
 \pdv{\Delta b_k}{\lambda} =  \sum_{l=1}^{l_{\mathrm{max}}} l \cdot A_l C^{l-1} \pdv{C}{\lambda} = \sum_{l=1}^{l_{\mathrm{max}}} B_l(D) C^{l-1},
\end{equation}
where \( D = \pdv{C}{l} \) and \( B_l(D) = l A_l D  \).
Now, we treat \( C \) and \( D \) as two independent variables and tune
their values in two passes: firstly the value of \( C \) is renormalized to
reproduce the \( \Delta b_2 \); and then we plug the resulting \( C \) in
Eq.~(\ref{eq:29}) and tune \( D \) to produce the expected second-order
result as given in Eq.~(\ref{eq:27}).

\begin{figure}[h]
  \centering
  \includegraphics[width=\linewidth]{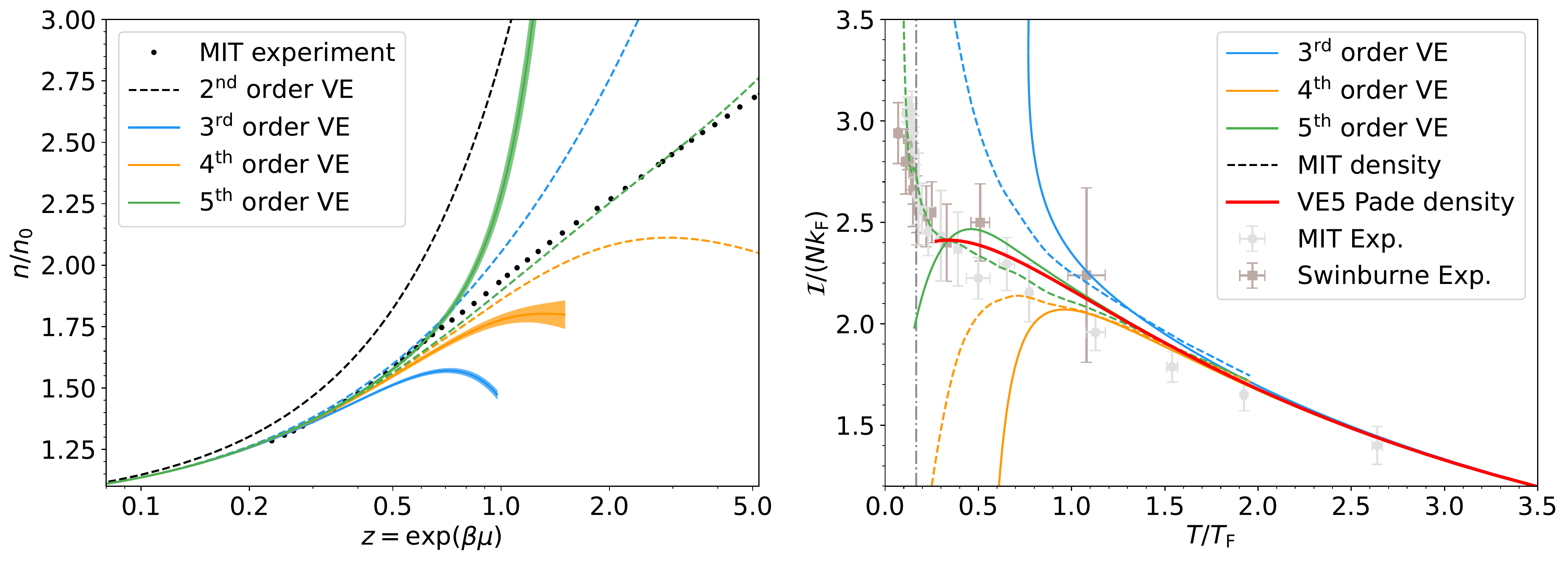}    
  \caption{{\bf Left}: Density \( n / n_0 \) as a function of fugacity
    \( z \). The colorful solid lines are the results using truncated VE
    at third (blue), fourth (orange) and fifth (green) order. The black
    dashed line is the result using second-order VE. The dashed lines are
    the results using Pad\'e resummation at order [a/b] The black dotted
    line is the experimental measurement by Ref.~\cite{MITExpKu2012S}. {\bf
      Right}: Dimensionless contact \( \mathcal{I} / (N k_F) \) as a function of
    reduced temperature \( T / T_F \). The blue, orange and green solid
    lines are the results using truncated VE for both the contact and the
    density, which is connected to the reduced temperature \( T / T_F
    \). The same color code is used. Colorful dotted lines are the results
    using the experimental density measurement and VE contact. The red
    curves are the results using resummation: the dotted is with
    experimental density and the solid with Pad\'e resummed density. The
    light gray and brown points are experimental determinations from
    Ref.~\cite{MITExpMukherjee2019PRL} and Ref.~\cite{ValeExpCarcy2019PRL}
    respectively.}
  \label{fig:homo-obs}
\end{figure}

In Fig.~\ref{fig:homo-obs}, we plot the density (left panel) and Tan's
contact (right panel) in the homogeneous case.
On the left panel, the solid lines (with uncertainty bands) are the results 
of using the VE at face value, i.e. truncated at a given order: blue for 
third order, red for fourth order, and green for 
fifth order. In each case, a corresponding dashed line of the same color shows
the results of Pad\'e resummation at that order.
Finally, the second-order results are shown as a black dashed line for
reference, while the black dots come from the MIT experiment of Ref.~\cite{MITExpKu2012S}.
For both the truncated VE and the resummed results, we found improved
agreement as higher-order contributions were included, the effect being most
notable when the resummation is applied. This inspires an open question under
investigation on the effect of the resummation when using even higher orders.
This is particularly intriguing as access to higher-order coefficients would allow 
more freedom in exploring a range of resummation techniques.

In the right panel of Fig.~\ref{fig:homo-obs} we show the dimensionless contact
in the form
\begin{equation}
  \label{eq:30}
  \frac{\mathcal{I}}{N k_F} = 3 \pi^2 \left(\frac{T}{T_F}\right)^2 \sum_{k=2} c_k z^k,
\end{equation}
as a function of the reduced temperature \( T / T_F \), which is related to
the density via
\begin{equation}
  \label{eq:31}
  \left(\frac{T}{T_F}\right)^3 = \left( \frac{8}{3 \sqrt{\pi} \lambda_T^3 n}\right)^{2}.
\end{equation}
Therefore, the resulting curve is an interplay between two independent series for the
density and the contact. In particular, the curve will be more sensitive to
the density estimate as both quantities implicitly depend on it. In
other words, in the region where we observed deviation in the left panel
between our results and experimental determinations, we expect errors in
both horizontal and vertical directions. This approximately corresponds to
\( z \gtrapprox 0.75 \), or \( T / T_F \lessapprox 1.4 \), for the truncated VE results.
At higher temperature, where our results coincide with the
experimental measurement, the main source of error is the expansion series for
\( \mathcal{I} \).

To better show \( \mathcal{I} \) as given in Eq.~(\ref{eq:24}), and isolate the
influence on the density, we use the experimental values
\cite{MITExpMukherjee2019PRL} to calculate the reduced temperature
\( T / T_F \) (thus providing an essentially exact density) and evaluate
the series \( \sum_{k=2}^{\infty} c_k z^k \) using Pad\'e resummation at different
orders (dashed lines). Overall, we observed the same trend as for the
density where the results are improved as higher-order contributions are
included.
Although the resummed fifth-order VE (red dashed line) seems to give 
very good results below the critical temperature, one should take such an
agreement with a grain of salt, as the error in the dimensionless contact is greatly
reduced by a factor of \( (T / T_F)^2 \). Furthermore, this result is free from errors in 
the density because we have used the experimental data for the latter.

To represent the more practical situation where we have no access to accurate
experimental measurement of the density, we use the best density estimate
(green dashed line in the left panel) and the final result is plotted as
the thick red solid line.
Even though the approximation begins to fail when approaching the
critical temperature, we find impressive agreement until
\( T / T_F \approx 0.4 \), roughly corresponding to the diverging point between
the Pad\'e and experimental results at \( z = 5.0 \) for the density.

\subsection{One-body operator example: the momentum distribution}
\label{sec:moment-distr}

Besides the density and pressure equations of state, the next simplest
quantity one can calculate is the expectation value of a one-body operator,
for instance the momentum distribution $\hat n_\sigma ({\bf p})$ of particles
with spin $\sigma$. Of course, here the only added difficulty is that the
operator singles out a specific momentum $\bf p$, but it warrants special
attention.

The starting point is the thermal expectation value
\beq
\langle \hat n_\sigma ({\bf p}) \rangle = 
\frac{1}{\mathcal Z} \Tr[e^{- \beta (\hat{H} - \mu \hat{N})} \hat n_\sigma ({\bf p})].
\eeq
Having available the virial expansion for $\mathcal Z$ already, we focus on the numerator
\beq
\label{eq:one-body-ve}
\Tr[e^{- \beta (\hat{H} - \mu \hat{N})} \hat n_\sigma ({\bf p})  ] = 
\sum_{N=1}^{\infty} \tr^{}_N \left [ e^{-\beta \hat{H}}  \hat n_\sigma ({\bf p})  \right] z^N,
\eeq
Thus, after expanding the denominator \( \mathcal{Z} \),
\beq
\langle \hat n_\sigma ({\bf p}) \rangle = \sum_{k=1}^{\infty} m_{\sigma,k} ({\bf p}) z^k,
\eeq
where
\bea
m_{\sigma,1}({\bf p}) &=& \tr^{}_1 \left [ e^{-\beta \hat{H}} \hat n_\sigma ({\bf p}) \right], \\
m_{\sigma,2}({\bf p}) &=& \tr^{}_2 \left [ e^{-\beta \hat{H}} \hat n_\sigma ({\bf p}) \right]
- Q_1 \tr^{}_1 \left [ e^{-\beta \hat{H}} \hat n_\sigma ({\bf p}) \right],\\
m_{\sigma,3}({\bf p}) &=& \tr^{}_3\left [ e^{-\beta \hat{H}} \hat n_\sigma ({\bf p}) \right] - Q_1 \tr^{}_2\left [ e^{-\beta \hat{H}} \hat n_\sigma ({\bf p}) \right]\\
& &+ \left( Q_1^2 - Q_2 \right) \tr^{}_1 \left [ e^{-\beta \hat{H}} \hat n_\sigma ({\bf p}) \right],
\eea
and so forth. (Note that, naturally, the content of $m_{\sigma,1}({\bf k})$ is entirely
noninteracting, as it corresponds to a single-particle subspace.)

The new ``virial coefficients'' \( m_{\sigma, k}({\bf p}) \) requires the
calculation of the Hilbert trace
\( \tr_N [e^{-\beta \hat{H}} \hat{n}_{\sigma}({\bf p})] \), denoted as
\( W_N[\hat{n}_{\sigma}({\bf p})] \) or just \( W_{N, \sigma} \) for shorthand. In
our method, its evaluation shares the same formalism as the usual partition
function \( Q_N \) as in Eqs.~\eqref{eq:12} and \eqref{eq:13}, i.e. the trace
\( W_{N, \sigma} \) encodes the particle statistics. Taking the
\( (2,1) \)-system for example,
\begin{equation}
  \label{eq:32}
  \begin{aligned}
    W^\text{F}_{21, \sigma} & = \tr^{}_\text{F}\left[ \mathcal M_{21}^{N_\tau} \mathcal{N}_{21, \sigma}(\mathbf{p}) \right]
    = \frac{1}{2!} \sum_{a b c} \left\{ \left[ \mathcal M_{21}^{N_\tau}  \mathcal{N}_{21, \sigma} (\mathbf{p})  \right]_{abc, abc} - \left[ \mathcal M_{21}^{N_\tau}  \mathcal{N}_{21, \sigma}(\mathbf{p}) \right]_{abc, bac}\right\}, \\
    W^\text{B}_{21, \sigma} & = \tr^{}_\text{B}\left[ \mathcal M_{21}^{N_\tau}  \mathcal{N}_{21, \sigma}(\mathbf{p}) \right]
    = \frac{1}{2!} \sum_{a b c} \left\{ \left[ \mathcal M_{21}^{N_\tau}  \mathcal{N}_{21, \sigma}(\mathbf{p})
      \right]_{abc, abc} + \left[ \mathcal M_{21}^{N_\tau}  \mathcal{N}_{21, \sigma}(\mathbf{p}) \right]_{abc,
        bac}\right\}.
  \end{aligned}
\end{equation}
where \( \mathcal{N}_{\sigma, \mathbf{p}} \) is the matrix elements of momentum density operator
\begin{equation}
  \label{eq:33}
  \begin{aligned}
    \mathcal{N}_{21, \sigma}(\mathbf{p}) & = \mel{\mathbf{p}_1 \mathbf{p}_2 \mathbf{p}_3}{\hat{n}_{\sigma}(\mathbf{p})}{\mathbf{q}_1 \mathbf{q}_2 \mathbf{q}_3}\\
    & = \left[ \delta_{\mathbf{p} \mathbf{p}_1} + \delta_{\mathbf{p} \mathbf{p}_2} \right] \braket{\mathbf{p}_1 \mathbf{p}_2 \mathbf{p}_3}{\mathbf{q}_1 \mathbf{q}_2 \mathbf{q}_3}\\
    & = \left[ \delta_{\mathbf{p} \mathbf{p}_1} + \delta_{\mathbf{p} \mathbf{p}_2} \right] \delta_{\mathbf{p}_1  \mathbf{q}_1} \delta_{\mathbf{p}_2  \mathbf{q}_2} \delta_{\mathbf{p}_3  \mathbf{q}_3}.
  \end{aligned}
\end{equation}

At leading-order up to the third order in fugacity, we obtain the change of momentum distribution with respect to that in the non-interacting case,
\begin{equation}
  \label{eq:37}
  \Delta n_{\sigma}(\mathbf{p}) = \frac{C}{\lambda_T^d} \exp(- \frac{d}{4 \pi} \tilde{k}^2) z^2 - 4 \frac{C}{\lambda_T^d} \exp(- \frac{d}{2 \pi} \tilde{k}^2) z^3,
\end{equation}
where \( \tilde{k} = \lambda_T k \) is the dimensionless momentum.
As in previous cases, we emphasize that this is merely an approximation at the lowest
non-trivial order in $C$, but it provides a qualitative guide and a reasonable
answer at weak coupling.

\subsection{Two-body operators and real-time evolution}
\label{sec:dens-dens-corr}

An advantage of our automated algebra method is its versatility. The formalism and
implementation can be adapted to more complicated observables without
introducing significant new complexity. In this section, we present the first
steps for a few such examples, which show the way for future work.

As a natural extension of the case of one-body operators,
Eq.~\eqref{eq:one-body-ve} is generalized to two-body operators:
\begin{equation}
  \label{eq:34}
  \begin{aligned}
    \expval{\hat{O}_1 \hat{O}_2} & = \frac{1}{\mathcal{Z}} \Tr[e^{-\beta (\hat{H} - \mu \hat{N})} \hat{O}_1 \hat{O}_2] \\
    & = \frac{1}{\mathcal{Z}} \sum_{N=1}^{\infty} \tr_N[ e^{-\beta \hat{H}} \hat{O}_1 \hat{O}_2 ] z^N.
  \end{aligned}
\end{equation}
From this point on, the evaluation of the $N$-particle Hilbert space trace
\( W(\hat{O}_1, \hat{O}_2) = \tr_N[ e^{-\beta \hat{H}} \hat{O}_1 \hat{O}_2] \)
follows the same methodology explained in the previous section.

An interesting two-body quantity is the density-density
correlation \( \expval{\hat{n}({\bf r}) \hat{n}(0) }\),
which is the central ingredient of the \emph{static structure factor}:
\beq
S(q) = \int d^3 r e^{-i {\bf q} \cdot {\bf r}} \expval{\delta \hat{n}({\bf r}) \delta \hat{n}(0) }
\eeq
where $\delta \hat X = \hat X - \expval{\hat X}$.
The latter approximately describes
the in-medium scattering rate in certain contexts~\cite{HOROWITZ2006326}. 
Recent studies~\cite{Alexandru2020PRLStructure,Alexandru2020PRC} 
examined this quantity within the VE up to second order, and it is interesting to
examine the effect of higher-order contributions.
Yet another example that has been intensely investigated recently is the viscosity. 
Specifically, in the work of Refs.~\cite{ViscosityEnss,ViscosityNishida,ViscosityHofmann}, 
the VE was applied to the calculation of both bulk and shear viscosity, which requires the
commutator of the contact operator. 

Our formalism can also be generalized beyond equilibrium systems by
including real-time evolution. The recent work of Ref.~\cite{QuenchSun2020PRL}
investigated the effects of an interaction quench in a bosonic system,
in an attempt to explain the results of the experiment of Ref.~\cite{EigenNature2018}.
Here, the system starts as noninteracting and in thermal equilibrium, and then an interparticle interaction is suddenly switched on. The process was investigated in 
Ref.~\cite{QuenchSun2020PRL} using the VE of the momentum distribution $n(k)$ 
up to second order and varying behaviors at 
low and high momenta $k$ were found. Although this is an interesting result, 
the effect of higher-order contributions remain an open problem and may not be small, 
which further motivates the extension of our framework for more general dynamic processes.

The quench process mentioned above is described by the Hamiltonian
\begin{equation}
  \label{eq:35}
  \hat{H} = \hat{H}_0 + \Theta(t)  \hat{V},
\end{equation}
where \( \hat{H}_0 \) is the free Hamiltonian and \( \hat{V} \) is the interaction, which is 
turned on at \( t = 0 \).
To study the evolution of a given operator $\hat O$ one starts with expressions similar to Eq.~(\ref{eq:34}), with the exception that the Hilbert space trace now becomes 
time-dependent
\begin{equation}
  \label{eq:36}
  W_N(\hat{O}, t) = \tr_N(e^{-\beta \hat{H_0}} e^{i t \hat{H}} \hat{O} e^{- i t \hat{H}}).
\end{equation}
Work in investigating these kinds of expressions with automated algebra,
specifically for the interaction quench, is underway. 

\section{Summary and conclusions}

The VE is an expansion of the quantum many-body thermodynamics in powers of the fugacity $z$ that is
capable of non-perturbatively characterizing such many-body systems in dilute, high-temperature regimes. The challenge of the VE is that calculating its coefficients $b_k$
at $n$-th order in the expansion requires analyzing the quantum $n$-body problem in some 
detail.

In this brief review, we have outlined some of the main methods used to
calculate $b_k$ and advocated an approach based on discretizing imaginary time, thus
obtaining a factorized transfer matrix, and 
automating the resulting algebra to calculate the trace of the $N_\tau$-th power of the transfer matrix. We have shown in detail how such a method is structured and how it is able to provide reliable results for up to $b_5$ for strongly coupled nonrelativistic fermions, focusing on the unitary limit. The automated algebra algorithm is fully parallelizable and may thus be extended beyond $b_5$ with increased computational power. 

We have
also shown how to generalize the approach to account for harmonic trapping potentials, 
neutron matter and attractive Bose gases (which require three-body forces). Furthermore,
the method is also straightforwardly generalized to observables other than the pressure,
such as the Tan contact, the momentum distribution, and the structure factor.
\\

\acknowledgments
{This material is based upon work supported by the National Science Foundation under 
Grants No. PHY1452635 and No. PHY2013078.}

\clearpage
\end{paracol}  
\newpage
\reftitle{References}


\externalbibliography{yes}

\bibliography{mdpireferences}

\end{document}